%% file: proc_lattice08_scholz.tex
\let\pdfoutput\defined

\documentclass{PoS}
\usepackage{amsmath, amsfonts, bbm, dsfont, mathrsfs,amssymb}
\usepackage{dcolumn}

\title{%
\vspace*{-1cm}
\begin{minipage}{\textwidth}
\begin{flushright}
\texttt{\footnotesize
PoS(LATTICE2008)095\\%
}
\end{flushright}
\end{minipage}\\[15pt]
Physical results from 2+1 flavor Domain Wall QCD%
}

\ShortTitle{Physical results from 2+1 flavor Domain Wall QCD}

\author{RBC and UKQCD Collaborations}

\author{\speaker{Enno E. Scholz}%
        \\
        Physics Department, Brookhaven National Laboratory, Upton, NY 11973, USA\\
        E-mail: \email{scholzee@quark.phy.bnl.gov}}

\input{abstract}

\FullConference{The XXVI International Symposium on Lattice Field Theory \\
		 July 14 - 19, 2008\\
		 Williamsburg, Virginia, USA}

\input{defin.tex}
\newlength{\closercaption}
\newlength{\afterTable}
\newlength{\afterFigure}
\newlength{\closersection} 
\begin{document}

\input{intro}

\input{chptfits_24c}

\input{chptfits_32c}

\input{conclusion}

\bibliography{references}

\bibliographystyle{JHEP-2} 

\end{document}

%% file: abstract.tex
\abstract{%
We review recent results for the chiral behavior of meson masses and decay constants and the determination of the light quark masses by the RBC and UKQCD collaborations. We find that one-loop SU(2) chiral perturbation theory represents the
behavior of our lattice data better than one-loop SU(3) chiral perturbation theory in both the pion and kaon sectors.
 
The simulations have been performed using the Iwasaki gauge action at two different lattice spacings with the physical spatial volume held approximately fixed at $(2.7\,{\rm fm})^3$. The Domain Wall fermion formulation was used for the 2+1 dynamical quark flavors: two (mass degenerate) light flavors with masses as light as roughly 1/5 the mass of the physical strange quark mass and one heavier quark flavor at approximately the value of the physical strange quark mass.

On the ensembles generated with the coarser lattice spacing, we obtain for the physical average up- and down-quark and strange quark masses $m_{ud}^{\overline{\rm MS}}(2\,{\rm GeV})=3.72\errRen{0.16}{0.33}{0.18}\,{\rm MeV}$ and $m_s^{\overline{\rm MS}}(2\,{\rm GeV})=107.3\errRen{4.4}{9.7}{4.9}\,{\rm MeV}$, respectively, while we find for the pion and kaon decay constants $f_\pi=124.1\err{3.6}{6.9}\,{\rm MeV}$, $f_K=149.6\err{3.6}{6.3}\,{\rm MeV}$. The analysis for the finer lattice spacing has not been fully completed yet, but we already present some first (preliminary) results.}

%% file: defin.tex
\newcommand{\err}[2]{(#1)_{\rm stat}(#2)_{\rm syst}}
\newcommand{\errRen}[3]{(#1)_{\rm stat}(#2)_{\rm ren}(#3)_{\rm syst}}

\def\nicefrac#1#2{\leavevmode\kern.1em\raise.5ex\hbox{\the\scriptfont0 #1}\kern-.1em/\kern-.15em\lower.25ex\hbox{\the\scriptfont0 #2}}

\newcommand{\SU}{{\rm SU}}

\newcolumntype{C}{>{$}c<{$}}
\newcolumntype{R}{>{$}r<{$}}
\newcolumntype{L}{>{$}l<{$}}

%% file: intro.tex
\section{Introduction}
\vspace*{\closersection}

Due to computer and algorithmic constraints we are not able to simulate
directly at the physical light quark mass. This necessitates performing
a chiral extrapolation. There are various ways that this extrapolation
can be done. We found that applying SU(2) partially quenched chiral perturbation theory (PQChPT) is working more reliable at next-to-leading order (NLO) compared to SU(3) PQChPT \cite{Lin:2007pt,Allton:2008pn}. The reason is that the strange quark mass is already too heavy to be described by the NLO terms in SU(3) ChPT. To be able to also extract quantities from the kaon sector, we introduced the SU(2) ChPT for kaon physics in \cite{Lin:2007pt,Allton:2008pn}. Recently other collaborations made similar observations about the limitations of NLO-SU(3) ChPT and also successfully applied (kaon) SU(2) ChPT in their analyses, e.g.\ \cite{Aoki:2008sm}.

We simulated QCD using $N_f=2+1$ flavors of Domain Wall fermions. Currently the mass of the heavy single flavor $m_h$ is kept fixed at a value close to the physical strange quark mass. We generated ensembles at multiple values for the mass $m_l$ of the two (degenerate) light quark flavors. 

Here we will focus on the extraction of the light quark masses, the pion and kaon decay constants and the low energy constants (LECs) of the SU(2) chiral Lagrangian. For a discussion of the treatment of the kaon bag parameter we refer to \cite{Allton:2008pn,Antonio:2007pb} and \cite{Kelly:proc} for recent developments. The remainder is organized as follows: in Sec.~\ref{sec:24c} we briefly describe our method to extract the physical results and estimate the systematic error and quote the results obtained at the ensembles with a lattice cut-off $1/a=1.73\,{\rm GeV}$. Before we conclude, we briefly present preliminary results obtained at a finer lattice spacing in Sec.~\ref{sec:32c}. 

For any unexplained notation and further details, we refer to \cite{Allton:2008pn}; especially App.~A therein contains an overview of the conventions followed here as well.

%% file: chptfits_24c.tex
\section{Physical results at $1/a=1.73\,{\rm GeV}$}
\label{sec:24c}
\vspace*{\closersection}

To obtain physical results on the $24^3\times64$, $L_s=16$ lattices (generated using the Iwasaki gauge action at $\beta=2.13$), we only used the ensembles with the two lightest dynamical light quark masses, $m_l=0.005$ and 0.01, which correspond to pion masses of 331 and 419 MeV, respectively. In the subsequent analysis, partially quenched (valence) masses $m_{x,y} \in \{ 0.001, 0.005, 0.01, 0.02, 0.03,\linebreak 0.04 \}$ have been used as well. The lattice scale $1/a=1.729(28)\,{\rm GeV}$ ($a=0.1141(18)\,{\rm fm}$), the physical average light and strange quark masses are fixed by the masses of the $\Omega^-$-baryon, the pion, and the kaon. In case of the $\Omega^-$-baryon this procedure includes an extrapolation in the dynamical light quark mass to the physical average up- and down-quark mass and a (valence) interpolation in the heavy dynamical mass to the point of the physical strange quark mass, cf.\ \cite{Allton:2008pn} for details. The residual mass parameter, measuring the remaining breaking of the chiral symmetry, turned out to be $m_{\rm res}=0.00315(2)$. In the following we will briefly describe our fit strategy and how the extrapolations in the pion and kaon sectors were performed and how the systematic errors were estimated.

\subsection{PQChPT fits}

\begin{figure}
\begin{center}
\includegraphics[angle=-90, width=.36\textwidth]{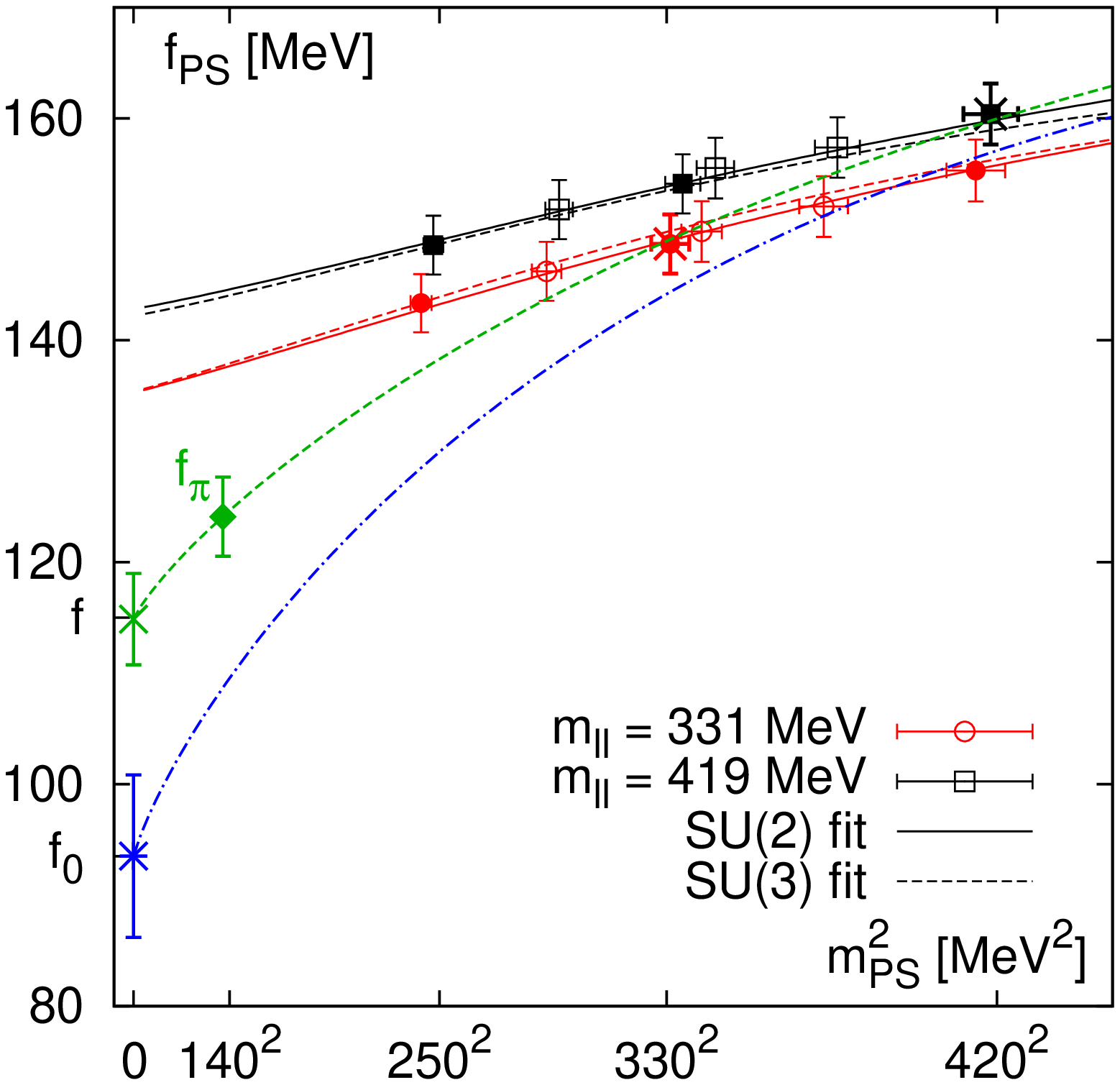}%
\includegraphics[angle=-90, width=.38\textwidth]{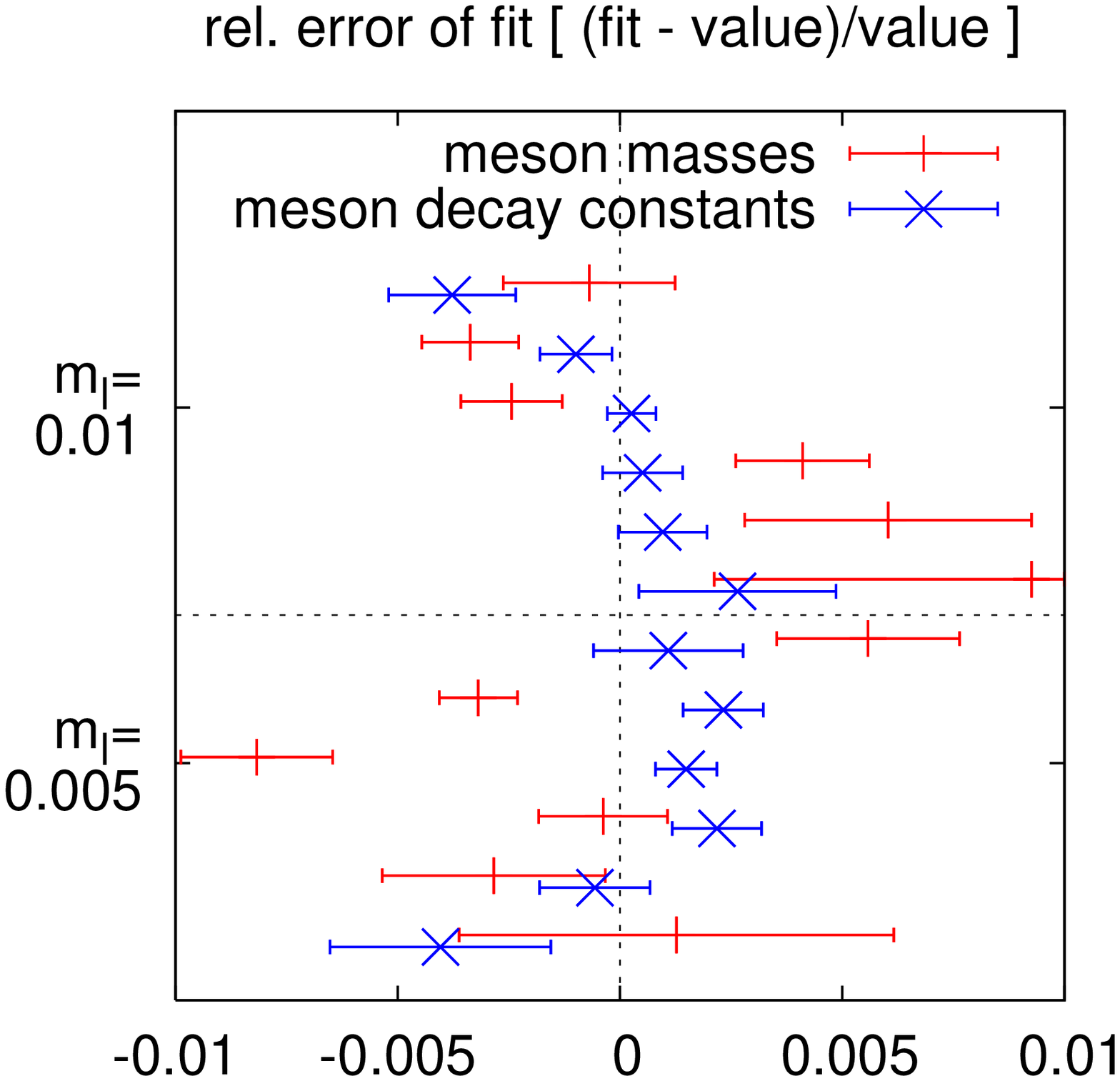}%
\end{center}
\vspace*{\closercaption}
\caption{\label{fig:Err_PR} \textit{Right panel:} Comparing the extrapolation to the SU(2) (\textit{dashed green curve}) and SU(3) (\textit{dashed-dotted blue curve}) chiral limit for the degenerate pseudoscalar decay constant. \textit{Left panel:} Relative deviation of the SU(2) PQChPT fit from the data.}
\vspace*{\afterFigure}
\end{figure}

As we already discussed extensively in \cite{Lin:2007pt,Allton:2008pn}, fitting to SU(3) NLO PQChPT including the physical strange quark mass is problematic. As shown for example in the left panel of Fig.~\ref{fig:Err_PR}, the decay constant receives large NLO-contributions (around 60--70\%) when extrapolated from pion masses in the range of 331--419 MeV to the SU(3) chiral limit ($f_0$). The decay constant in the SU(2) chiral limit $f$ (in which the strange quark mass is not sent to zero but kept fixed (close) to its physical value) receives a much smaller (30--40\%) NLO-contribution. Also we observed that applying PQChPT to data with meson masses in the region of the physical kaon mass, does not lead to reasonable fits if only terms up to NLO are considered. Therefore, we simultaneously fitted our data for the meson masses and decay constants to SU(2) NLO PQChPT imposing a cut on the average quark mass of $m_{\rm avg}\leq0.01$ (corresponding to $m_{PS}\leq 420\,{\rm MeV}$), see Fig.~10 from \cite{Allton:2008pn}. From the meson mass fit we are able to determine the value $m_{ud}=(m_u+m_d)/2$ for the physical average light quark mass. Finally, we extrapolated the meson decay constant to this point to predict $f_\pi$. We are aware that our data is correlated within the two ensembles (correlations between the different valence masses and between the meson masses and decay constants) but our statistics (for each ensemble 45 jackknife blocks made from 2 measurements) was not sufficient to obtain a reliable estimate of the (inverse) correlation matrix for the 2x6 data points per ensemble as needed in a correlated fit. For that reason, we refrained from using a correlated fit. From the uncorrelated (simultaneous) fit we obtained a $\chi^2/{\rm d.o.f.}$ of 0.3. As shown in the right panel of Fig.~\ref{fig:Err_PR}, the relative deviation of the fit from the data is always less than 1\%. Note, that we are not fitting to an exact theory, ChPT is an expansion around zero quark masses and higher orders (which were omitted here) are expected to account for those deviations.

The extrapolation to $m_{ud}$ in the kaon sector was done using kaon SU(2) as presented in \cite{Lin:2007pt,Allton:2008pn} and references therein. We did the extrapolation at two different (valence) masses for the heavy quark, $m_y=0.03$ and 0.04 and linearly interpolated between those. From the physical value of the (quadratically averaged) kaon mass we obtain the strange quark mass $m_s$ and then in turn the kaon decay constant $f_K$ at that point. Example plots are shown in Figs. 11 and 12 of \cite{Allton:2008pn}.

\subsection{Systematic errors}

We have to include estimates for the systematic errors due to the following four sources: finite volume of the simulated lattice box, the absence of a continuum extrapolation, corrections from higher orders in (PQ)ChPT, and the fact that our simulated heavy quark mass turned out to be roughly 15\% higher than the physical strange quark mass.

In case of the finite volume effects in our simulated $(2.74\,{\rm fm})^3$ box, we repeated the SU(2) fits this time including finite volume correction terms (see App.~C in \cite{Allton:2008pn} and references therein). We assigned the difference between those fits and our original fits as the FV-systematic error. Plots of the correction factor can be found in Fig.~13 of \cite{Allton:2008pn}. A comparison of our finite volume correction factors for our meson masses and decay constants at the dynamical points with the resummed method of \cite{Colangelo:2005gd} shows good agreement, see Tab.~\ref{tab:FVcorr}.

Since the analysis on the ensembles generated at a finer lattice spacing is not yet finished (for preliminary results see Sec.~\ref{sec:32c}) for the moment we estimate the effect from the missing continuum extrapolation to be 4\%, which corresponds to $(a\Lambda_{\rm QCD})^2$.

The higher order effects in (PQ)ChPT are taken into account as the difference between our original fits and fits using a larger cut-off in the average quark mass ($m_{\rm avg}\leq0.02$). Here we had to introduce analytic NNLO-terms to obtain a reasonable agreement between our data and the fits. Also, since with only a limited set of dynamical quark masses we could not include all possible analytic NNLO-terms, we conservatively doubled the difference to estimate the systematic error due to higher order terms in (PQ)ChPT.

With only one value for the dynamical heavy quark mass, an exploration of the effects due to shifting $m_h$ was not possible. Therefore, we had to rely on the predictions of SU(3) ChPT to estimate the size of the moderate (15\%) shift from $m_h$ to $m_s$. More details on the conversion from SU(3) LECs to those of SU(2) and how to obtain the ``$m_s\neq m_h$'' systematic error therefrom can be found in \cite{Allton:2008pn}. 

The final results given in the following subsection contain the systematic errors discussed above added in quadrature. Table XII of \cite{Allton:2008pn} gives a detailed breakdown of the total error into the different sources. In case of quantities which have to be renormalized in a certain scheme, we provide the renormalization error separately. (We usually quote results in the $\overline{\rm MS}$-scheme at 2 GeV, using the Rome-Southampton RI-MOM method. See \cite{Aoki:2007xm} and references therein.)

\subsection{Final results}

Including the (estimates of the) systematic errors discussed in the previous subsection, we quote the following physical results from our SU(2) (PQ)ChPT analysis at $1/a=1.73\,{\rm GeV}$:%
\begin{eqnarray*}
f_\pi=124.1\err{3.6}{6.9}\,{\rm MeV},\\
f_K=149.6\err{3.6}{6.3}\,{\rm MeV},&&f_K/f_\pi=1.205\err{0.018}{0.062},\\[5pt]
m_{ud}^{\overline{\rm MS}}(2\,{\rm GeV})=3.72\errRen{0.16}{0.33}{0.18}\,{\rm MeV},\\ 
m_s^{\overline{\rm MS}}(2\,{\rm GeV})=107.3\errRen{4.4}{9.7}{4.9}\,{\rm MeV},&&\tilde{m}_{ud}:\tilde{m}_s=1:28.8\err{0.4}{1.6}.
\end{eqnarray*}
Furthermore, the SU(2) LECs were determined as 
\begin{eqnarray*}
f=114.8\err{4.1}{8.1}\,{\rm MeV},&& B^{\overline{\rm MS}}(2\,{\rm GeV})=2.52\errRen{0.11}{0.23}{0.12}\,{\rm GeV},\\
\bar{l}_3=3.13\err{0.33}{0.24},&& \bar{l}_4=4.43\err{0.14}{0.77}.
\end{eqnarray*}
%
%

%% file: chptfits_32c.tex
\begin{table}
\begin{center}
\begin{tabular}{lc*{2}{|cc}}
\hline\hline
 &  & \multicolumn{2}{c|}{$R_m[\%]$} & \multicolumn{2}{c}{$-R_f[\%]$} \\
 & $m_{ll}[{\rm MeV}]$ & SU(2) & CDH & SU(2) & CDH \\\hline
%
$24^3$, $V\approx(2.74\,{\rm fm})^3$ & 331 & 0.09(.01) & 0.13(.03) & 0.36(.03) & 0.32(.00)  \\
 & 419 & 0.03(.00) & 0.04(.01) & 0.10(.01) & 0.09(.00) \\[3pt]
$32^3$, $V\approx(2.60\,{\rm fm})^3$ & 307 & 0.16(.01) & 0.26(.07) & 0.62(.03) & 0.64(.01) \\
 & 364 & 0.07(.01) & 0.12(.03) & 0.28(.01) & 0.28(.00) \\
 & 419 & 0.04(.00) & 0.06(.02) & 0.14(.01) & 0.13(.00) \\
\hline\hline
\end{tabular}
\end{center}
\vspace*{0.7\closercaption}
\caption{\label{tab:FVcorr}Finite volume correction factors obtained from our SU(2) PQChPT fits including FV-terms compared to results interpolated from \cite{Colangelo:2005gd} (\textit{CDH}).}
\vspace*{\afterFigure}
\end{table}

\section{First results at larger cut-off}
\label{sec:32c}
\vspace*{\closersection}

Currently, our collaborations are in the middle of finishing measurements on a second set of ensembles, generated at a finer lattice spacing. We simulated three different light quark masses $m_l=0.004$, 0.006, and 0.008 at a fixed heavy quark mass, $m_h=0.03$ on $32^3\times64$, $L_s=16$ lattices with the gauge coupling set to $\beta=2.25$ (Iwasaki gauge action). A first estimate of the lattice cut-off obtained from measuring the Sommer-parameter $r_0/a$ gives $1/a=2.42(4)\,{\rm GeV}$ ($a\approx0.08\,{\rm fm}$), where $r_0=0.47\,{\rm fm}$ has been assumed. The PQChPT fits will include valence masses $m_{x,y}\in\{0.002, 0.004, 0.006, 0.008, 0.025, 0.03\}$. Using the above lattice cut-off, our dynamical pion masses are 307, 366, and 418 MeV, respectively, whereas the lightest valence pion mass reaches 236 MeV. The preliminary value for the residual mass parameter is $6.76(0.11)\cdot10^{-4}$, i.e.\ almost by a factor of 5 smaller than on the coarser lattices used in the previous analysis.

Since we have not reached a sufficiently high statistics on the three ensembles, we will refrain from quoting any physical results from this analysis. The following subsections contain the preliminary fits to SU(2) PQChPT and also (unquenched) ChPT, since here we have enough data to even perform a fit just including dynamical data points.

\subsection{PQChPT fits}

\begin{figure}
\begin{center}
\begin{minipage}{.45\textwidth}
\begin{center}
\includegraphics[angle=-90, width=.9\textwidth]{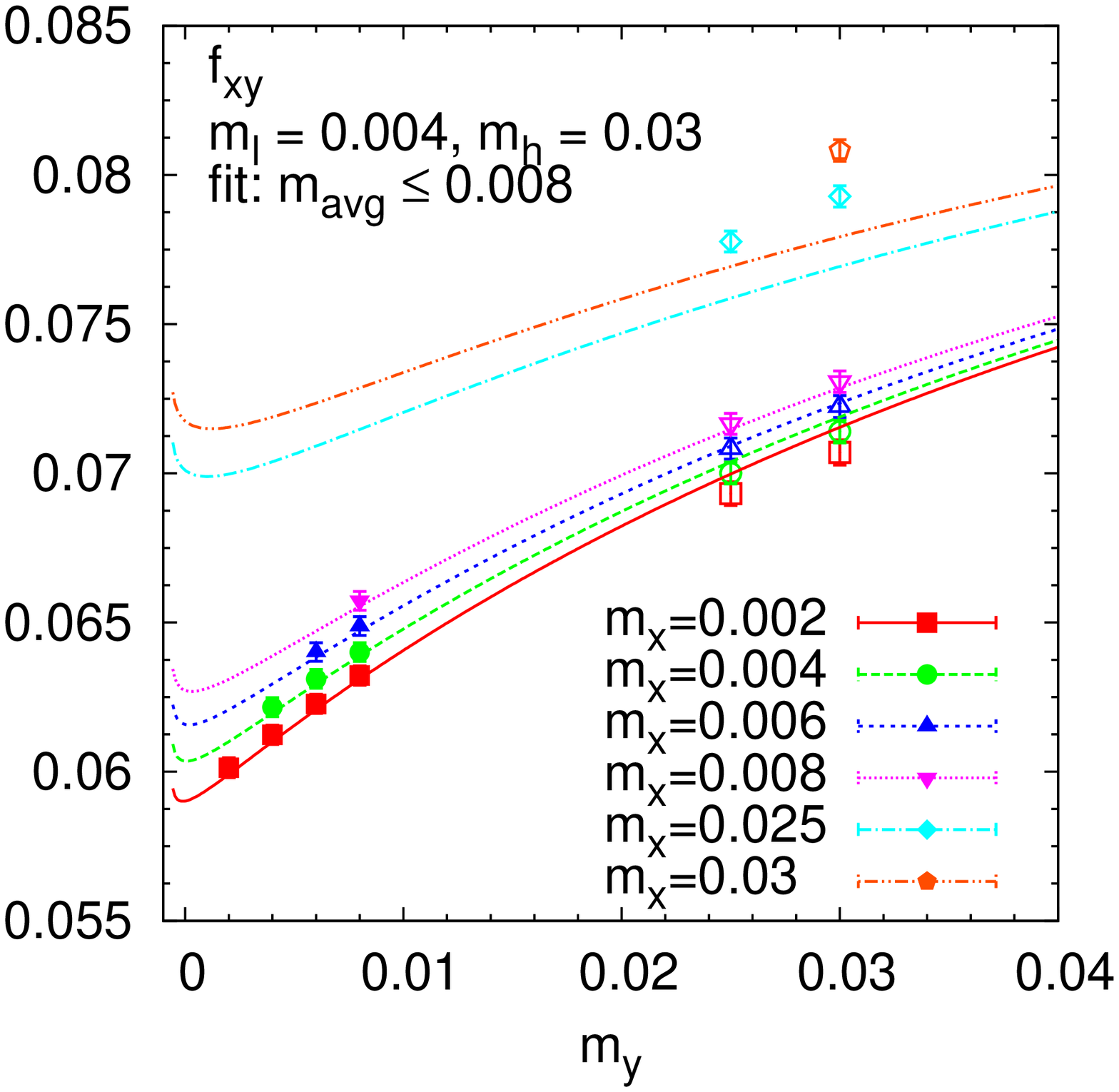}\\
\includegraphics[angle=-90, width=.9\textwidth]{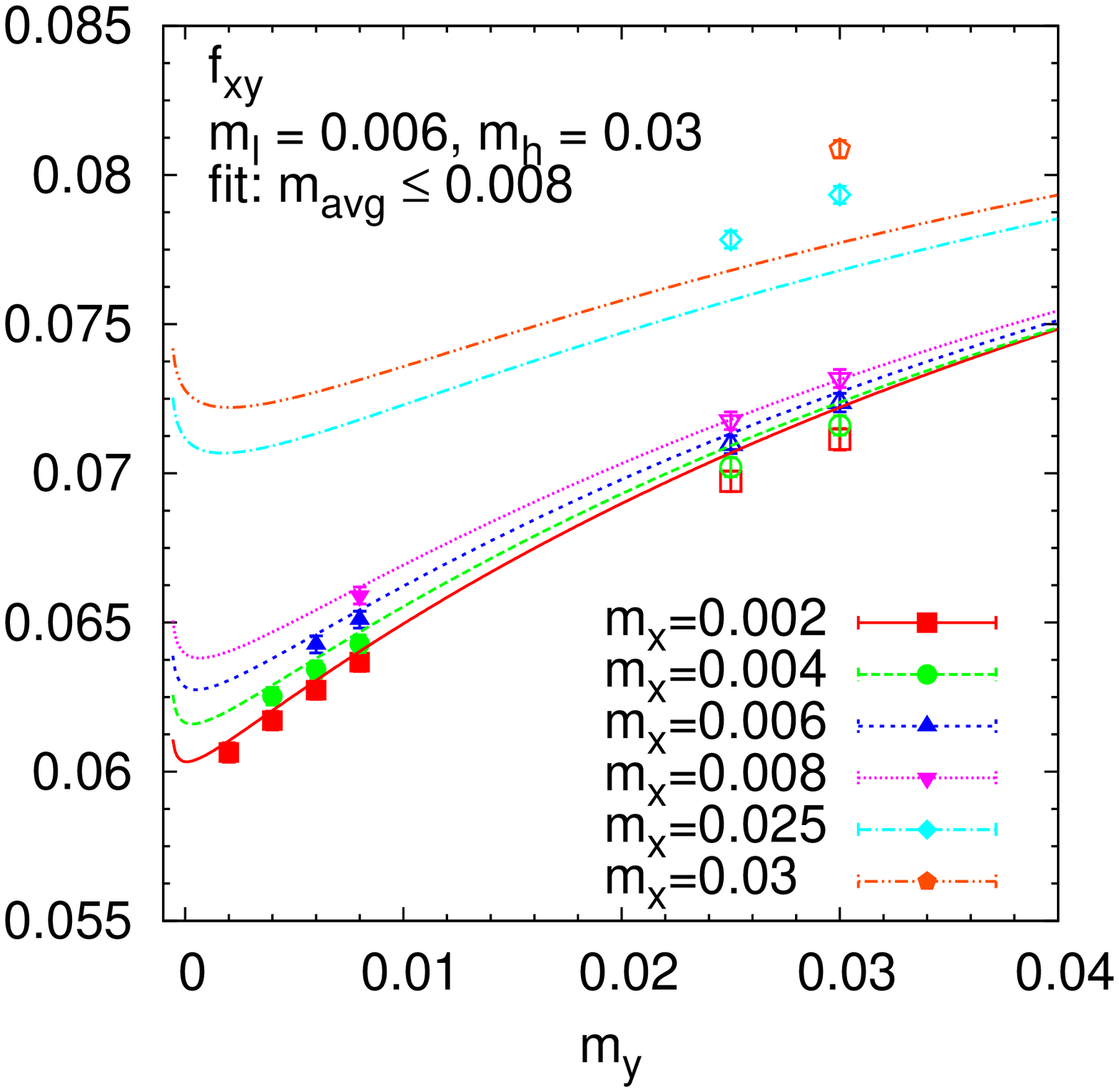}\\
\includegraphics[angle=-90, width=.9\textwidth]{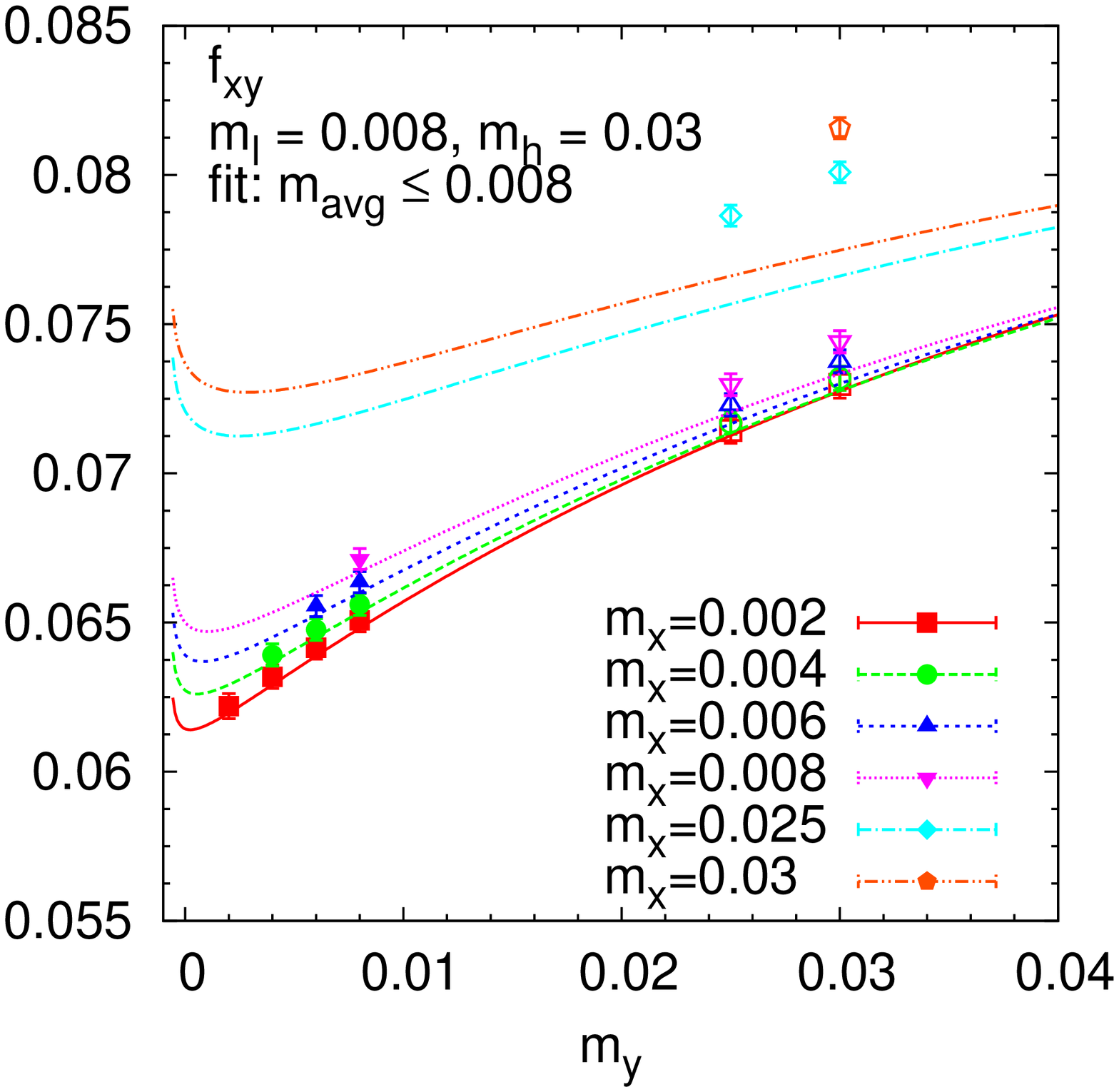}
\end{center}
\end{minipage}%
\begin{minipage}{.45\textwidth}
\begin{center}
\includegraphics[angle=-90, width=.87\textwidth]{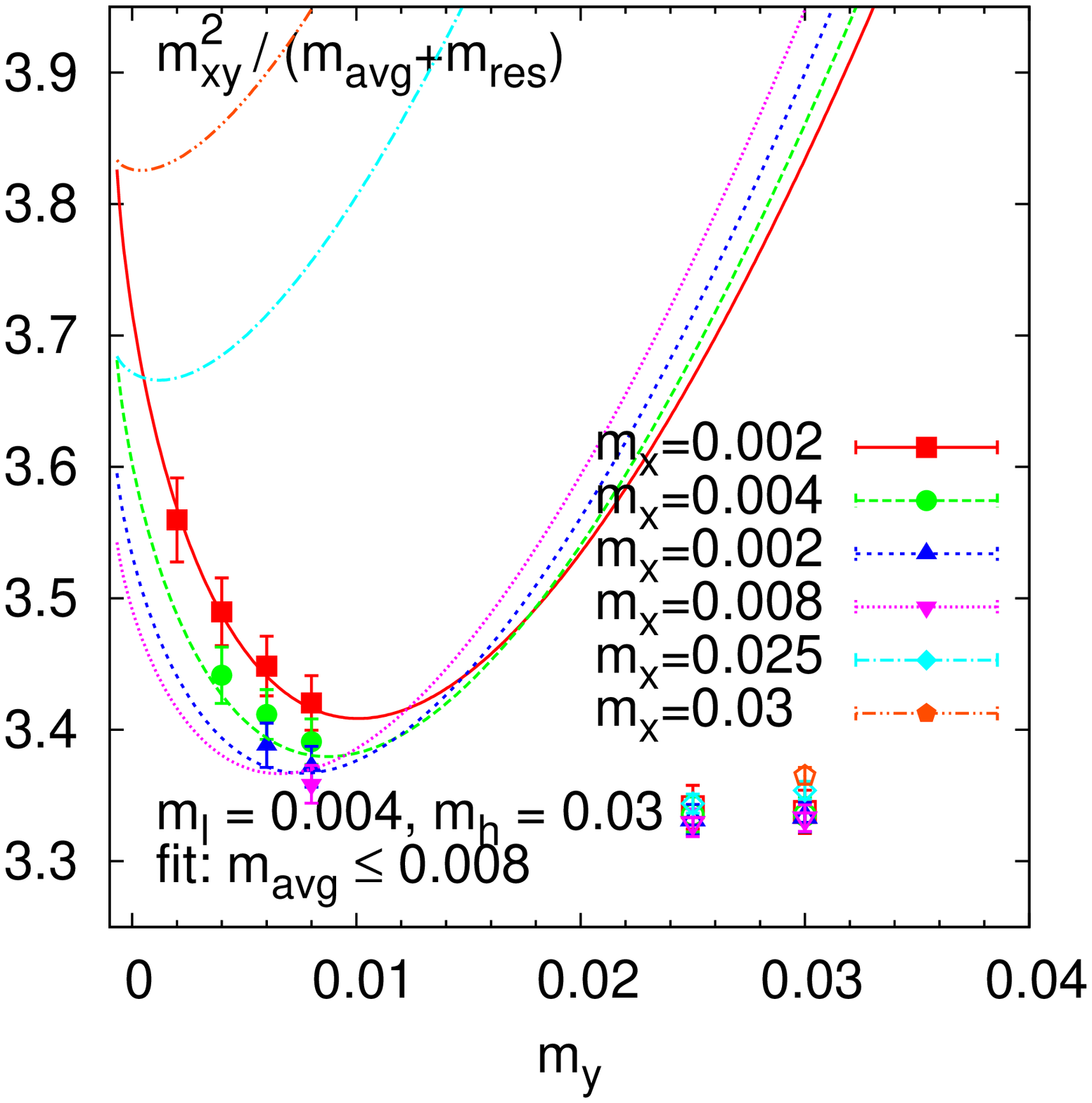}\\
\includegraphics[angle=-90, width=.87\textwidth]{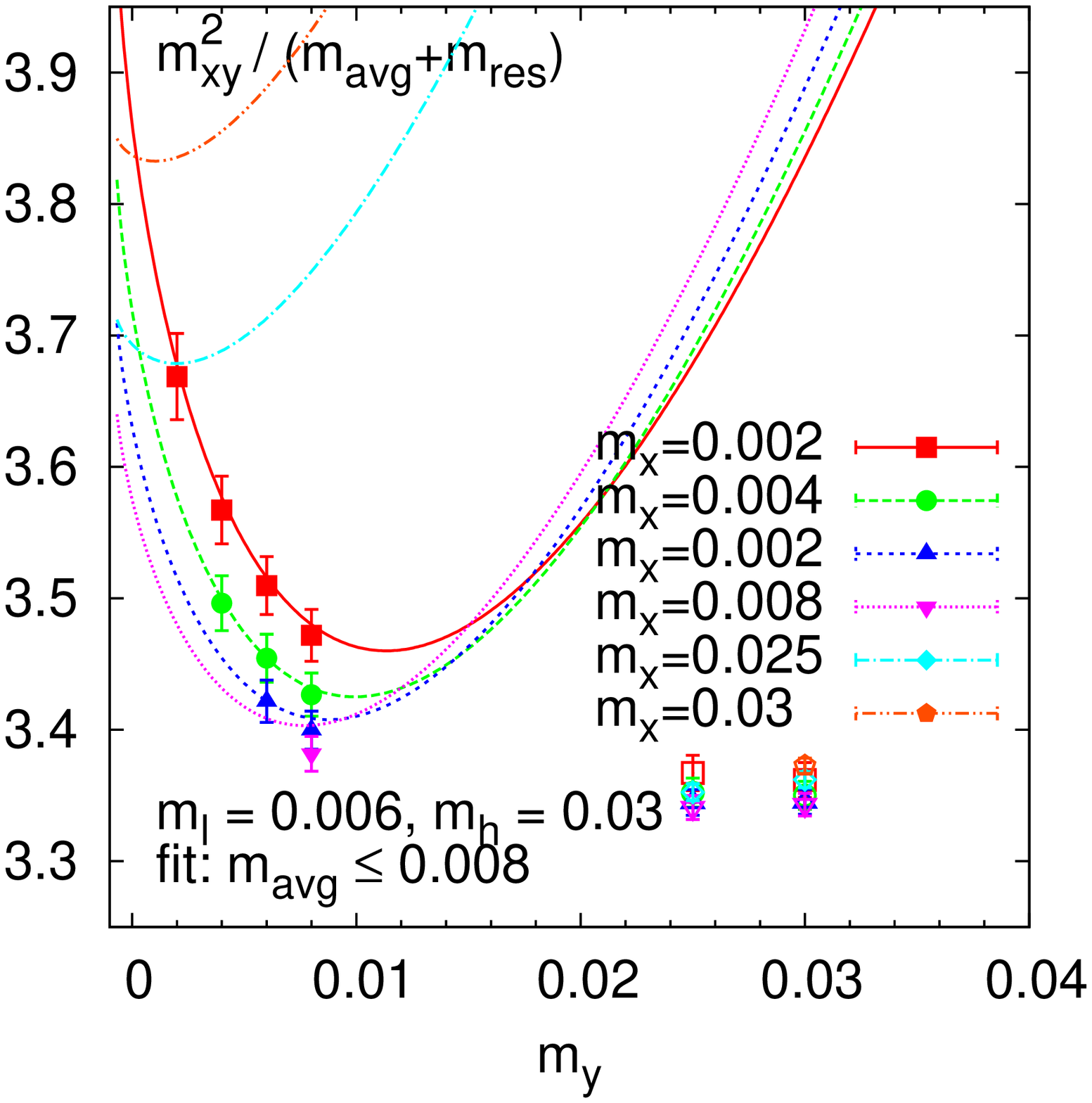}\\
\includegraphics[angle=-90, width=.87\textwidth]{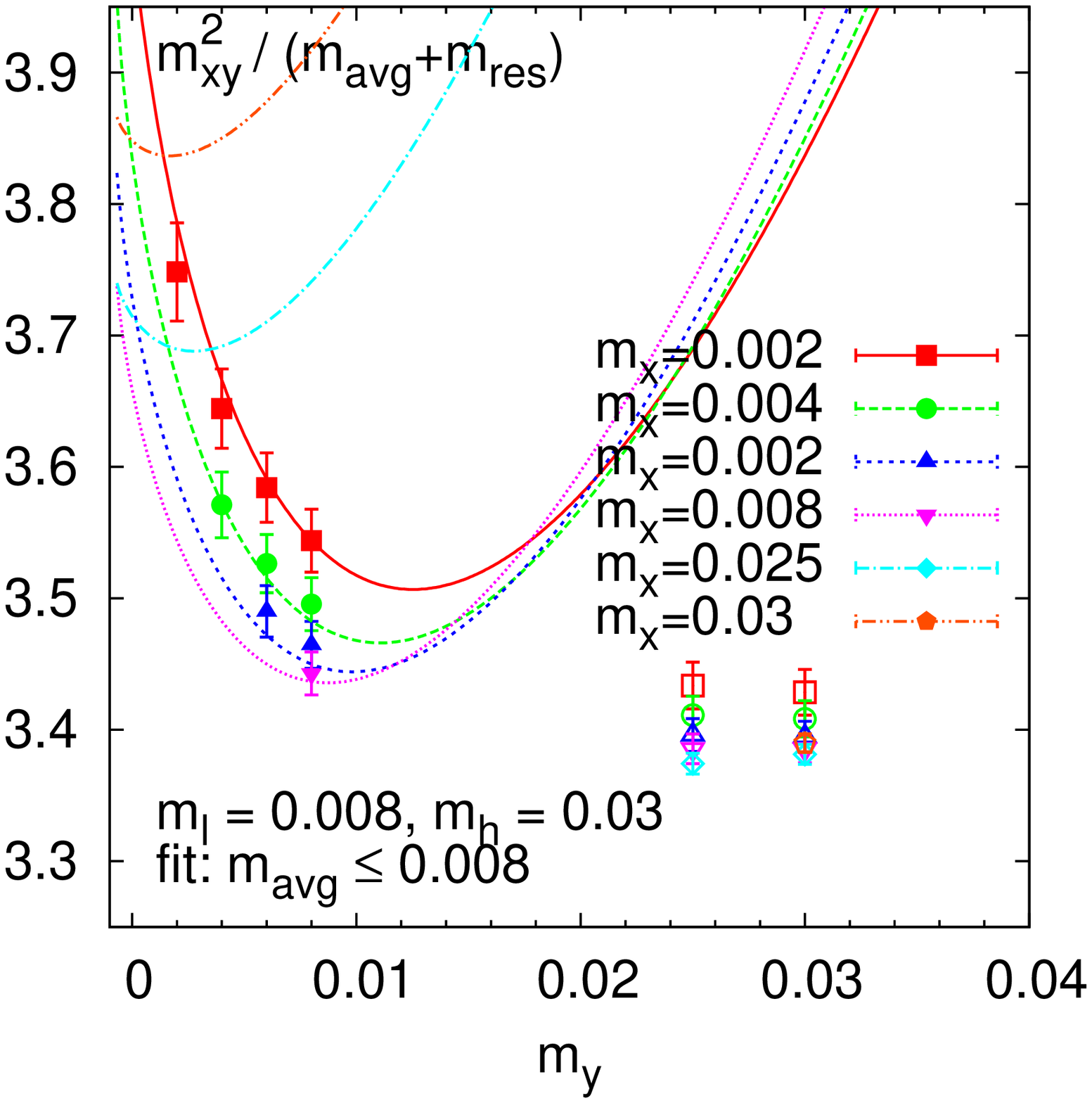}
\end{center}
\end{minipage}
\end{center}
\vspace*{\closercaption}
\caption{\label{fig:SU2_32c}Combined $\SU(2)\times \SU(2)$ fits for the meson decay constants \textit{(left panels)} and masses \textit{(right panels)} at three different values for the light sea quark mass ($32^3\!\!\times\!\!64\!\!\times\!\!16$ lattices), valence mass cut $m_{\rm avg}\leq0.008$. Points marked by \textit{filled symbols} were included in the fit, while those with \textit{open symbols} were excluded.}
\vspace*{\afterFigure}
\end{figure}

In Fig.~\ref{fig:SU2_32c} we show simultaneous (uncorrelated) fits of the meson decay constants and masses to NLO-PQChPT formulae, where a cut of $m_{\rm avg}\leq 0.008$ ($m_{PS}\leq 420\,{\rm MeV}$) in the average quark mass has been applied. The obtained $\chi^2/{\rm d.o.f.}$ of 0.6 is reasonable, although for some points the fit deviates as much as 1.0(0.7)\% from the data. But since here the statistical uncertainty of 0.7 percent-points is rather large, we will have to wait for the higher statistics to see if these deviations will disappear or remain.

Finite volume effects may also be of more importance in the analysis of the $32^3$ ensembles, since (given the preliminary number for the lattice cut-off quoted above) the spatial volume $V\approx (2.6\,{\rm fm})^3$ is slightly smaller compared to our $24^3$ ensembles. For the dynamical pion mass we still have $m_{ll}L\approx$ 4.1--5.5, whereas for our lightest valence pion mass, we only have $m_{xx}L\approx 3.1$. In Tab.~\ref{tab:FVcorr} we give the finite volume correction factors for our dynamical points as obtained from our SU(2) fits including finite volume terms and compare them to the results from the resummed L\"uscher formula of \cite{Colangelo:2005gd} and the results from the $24^3$ ensembles. The correction factors for our lightest valence meson ($m_x=m_y=0.002$) are $R_m=0.96(.04)\%$, $-R_f=1.00(.04)\%$ at $m_l=0.004$ and $R_m=2.00(.08)\%$, $-R_f=0.41(.02)\%$ at $m_l=0.008$.

\subsection{ChPT fit}

\begin{figure}
\begin{center}
\includegraphics[angle=-90, width=.36\textwidth]{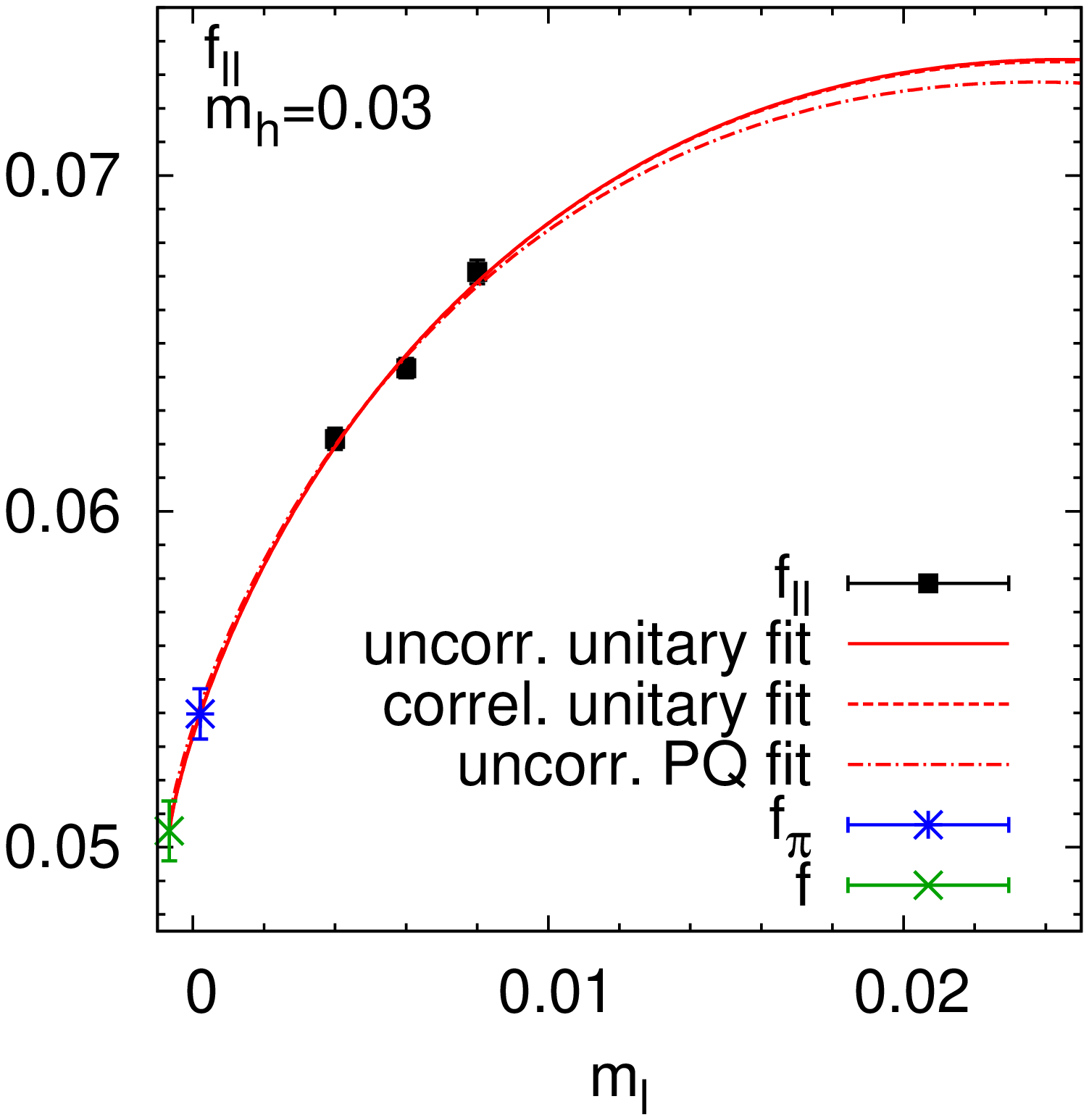}%
\includegraphics[angle=-90, width=.36\textwidth]{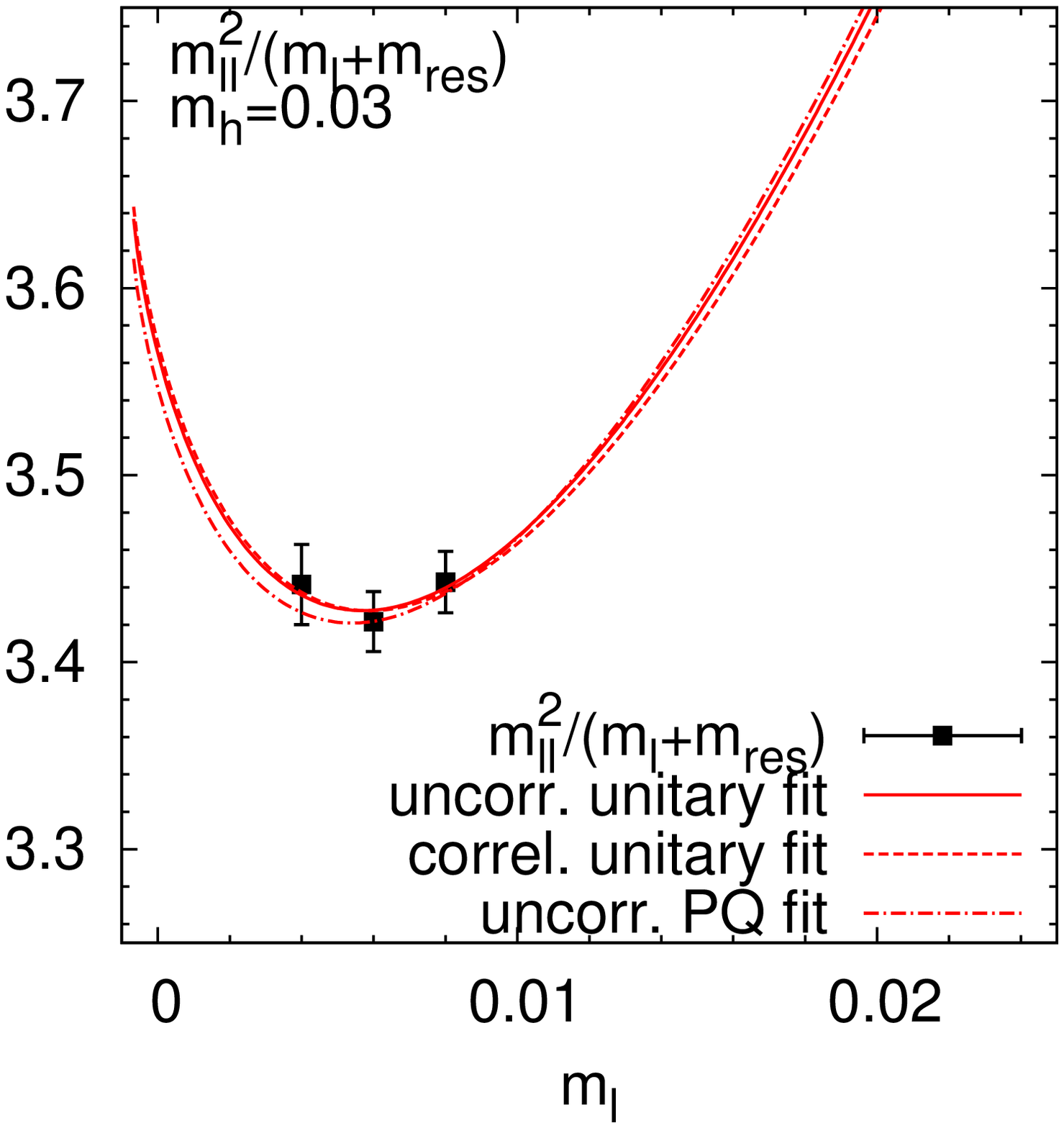}%
\end{center}
\vspace*{\closercaption}
\caption{\label{fig:SU2_32c_unit} Combined ChPT %
(\textit{solid} and \textit{dashed curves} for uncorrelated and correlated fit, respectively)
 and PQChPT fits %
(\textit{dashed-dotted curves})
, \textit{left panel:} meson decay constant, \textit{right panel:} meson mass.}
\vspace*{.65\afterFigure}
\end{figure}

Having three dynamical light quark masses which can be considered to be light enough to be described by NLO SU(2) ChPT, a combined fit just including those dynamical points becomes possible, too. In this case we have four fit parameters (the two LO-LECs: $f$ and $B$ plus two NLO-LECs: $l_3^r$, $l_4^r$) and six data points (meson decay constant and mass for each dynamical point). In Fig.~\ref{fig:SU2_32c_unit} we show the results of the combined (uncorrelated) fit (\textit{solid curves}), noting that the results for the fitted parameters are in good agreement with those obtained from the fit to the data including partially quenched points as well (\textit{dashed-dotted curves}). So we do not observe any artifacts  from partially quenching in our data.

Furthermore, since we now only have to deal with a 2x2 correlation matrix for each ensemble, we are also able to perform a correlated, combined fit (\textit{dashed curves} in Fig.~\ref{fig:SU2_32c_unit}) to our dynamical data, whose results are almost identical to those from the uncorrelated fit.  

%% file: conclusion.tex
\section{Conclusions and outlook}
\vspace*{\closersection}

The physical results obtained from the $24^3$ ensemble at $1/a=1.73\,{\rm GeV}$ demonstrate the successful application of SU(2) PQChPT. Currently the statistics on the two lightest ensembles used in that analysis is extended to further reduce the statistical uncertainty. With the three ensembles at a second, finer lattice spacing, we will be able to see the behavior in the continuum limit and improve our estimate of the systematic error associated with that missing extrapolation.

{\noindent\textbf{Acknowledgments.} I am thankful to all the members of the RBC and UKQCD Collaborations. The computations were done on the QCDOC machines at University of Edinburgh, Columbia University, and Brookhaven National Laboratory and the BG/P machines at Argonne National Laboratory available through the ESP and INCITE programs. The author was supported by the U.S.\ Dept.\ of Energy under contract DE-AC02-98CH10886.}